\documentstyle[prb,twocolumn,aps]{revtex}
\tighten
\begin{document}

\title{Effect of long-range Coulomb interaction 
on shot-noise suppression \\ in ballistic transport}
\author{T. Gonz\'alez, O. M. Bulashenko,\cite{byline} J. Mateos, and D. Pardo}
\address{
Departamento de F\'{\i}sica Aplicada, Universidad de Salamanca,
Plaza de la Merced s/n, E-37008 Salamanca, Spain}
\author{L. Reggiani}
\address{
Istituto Nazionale di Fisica della Materia, Dipartimento di Scienza
dei Materiali, Universit\`a di Lecce \\ Via Arnesano, 73100 Lecce, Italy}
\address{\rm (19 December 1996)}
\address{\rm cond-mat/9703110, Phys. Rev. B, v.56, 6424 (1997)}
\address{\parbox{14cm}{\rm \mbox{ }\mbox{ }\mbox{ }
We present a microscopic analysis of shot-noise suppression due to long-range
Coulomb interaction in semiconductor devices under ballistic transport
conditions. An ensemble Monte Carlo simulator self-consistently coupled with a
Poisson solver is used for the calculations. A wide range of injection-rate
densities leading to different degrees of suppression is investigated.
A sharp tendency of noise suppression at increasing injection densities is
found to
scale with a dimensionless Debye length related to the importance of
space-charge effects in the structure.
}}
\address{\mbox{ }}

\maketitle

The phenomenon of shot noise, associated with the randomness in the flux of
carriers crossing the active region of a device, has become a fundamental issue
in the study of electron transport through mesoscopic devices.
In particular, the possibility of shot-noise suppression has recently attracted
a lot of attention, both theoretically and experimentally. \cite{jong}
At low frequency (small compared to the inverse transit time through the active
region) the power spectral density of shot noise is given by 
$S_I=\gamma 2qI$, where $I$ is the dc current, $q$ is the electron charge, 
and $\gamma$ is the suppression factor. 
When the carriers crossing the active region are uncorrelated, 
full shot noise with $\gamma$=1 (Poisson statistics) is observed.
However, correlations between carriers can reduce the shot-noise value, 
giving $\gamma<1$.
In real mesoscopic devices different types of mechanisms resulting in shot-noise
suppression can be distinguished: (i) statistical correlations due to the Pauli
exclusion principle (important for degenerate materials obeying Fermi
statistics), (ii) short-range Coulomb interaction (electron-electron
scattering), and (iii) long-range Coulomb interaction (by means of the
self-consistent electric potential).
While the first two mechanisms have been extensively discussed in solid-state
literature, \cite{jong} the last one has received less attention, \cite{ziel}
although its role in shot-noise suppression has been known for a long time in
vacuum-tube devices. \cite{north}
The only exception that should be mentioned is the Coulomb blockade in
resonant-tunneling devices, which can be also referred to the last mechanism of
suppression.
The blockade is provided by a built-in charge inside a quantum well which
redistributes the chemical potential and prevents the incoming carriers 
from passing through the well, thereby resulting in carrier correlation and 
shot-noise suppression (see the experimental evidence \cite{birk}).
The Coulomb blockade is a consequence of long-range Coulomb interaction, and it
acts under the {\em sequential} tunneling regime of carrier transport.

The main objective of the present paper is to prove the importance of long-range
Coulomb interaction between the carriers on the shot-noise power spectrum
{\em under the ballistic regime of electron transport}.
The ballistic regime is now accessible in modern mesoscopic devices like
electron waveguides, quantum point contacts, etc., which have characteristic
lengths of the order, or smaller, than the carrier mean free path.
The current existing theories invoked to interpret the experimentally observed
shot-noise suppression in such devices \cite{li,reznikov,kumar} assume that
carriers move inside the device without inducing any redistribution of the
electric potential. We use a more rigorous approach which includes long-range
Coulomb interaction between the carriers by considering the carrier transport in
the {\em self-consistent} potential governed by the Poisson equation.
We show that under the ballistic regime this interaction is crucial, and the noise
characteristics are strongly modified depending on whether the carrier
correlation (mediated by the field) is taken into account or not.

To this purpose we consider a simple structure: a lightly doped active region of
a semiconductor device sandwiched between two heavily doped contacts injecting
the carriers into the active region. The device then acts similarly to a vacuum
diode, with a relevant difference in the fact that there are two
opposing currents
instead of a single current.
Electrons are emitted from the contacts according to a thermal-equilibrium
Maxwell-Boltzmann distribution, and they move ballistically inside the active
region (the mean free path is considered to be much larger than the distance $L$
between the contacts) according to the semiclassical equations of motion.
The fluctuating emission rate at the contacts is taken to follow a Poisson
statistics. This means that the time between two consecutive electron emissions
is generated according to the probability density $P(t)=\Gamma e^{-\Gamma t}$,
where $\Gamma=\frac{1}{2}n_c v_{\rm th}S$ is the injection-rate density, with
$n_c$ the electron density at the contact, $S$ the cross sectional area 
of the device, and $v_{\rm th}=\sqrt{2k_B T/(\pi m)}$ the thermal velocity 
($T$ is the lattice temperature, $k_B$ Boltzmann constant, and 
$m$ the electron effective mass).
The electron gas is assumed to be nondegenerate to exclude possible
correlations due to the Fermi statistics. For simplicity, the conduction band of
the semiconductor is considered to be spherically parabolic.
Both the time-averaged current and the current fluctuations inside the active
region of the device are analyzed for different bias voltages applied between
the contacts. The calculations are performed by using an ensemble Monte Carlo
simulator self-consistently coupled with a Poisson solver (PS).
By using this approach we can analyze much more general situations than those
studied in previous analytical calculations. \cite{ziel}

For the calculations we used the following set of parameters: $T$=300 K,
$m=0.25m_0$, dielectric constant $\varepsilon=11.7\varepsilon_0$, sample length
$L=2000\AA$, and contact doping $n_c$ ranging between $10^{13} {\rm cm}^{-3}$
and $4\times 10^{17} {\rm cm}^{-3}$ (always at least two orders of magnitude
higher than the sample doping). However, we must stress that the results we are
going to present do not depend on the particular values of these parameters but
only on the dimensionless length $\lambda=L/L_{Dc}$, where
$L_{Dc}=\sqrt{\varepsilon k_BT/q^2n_c}$ is the Debye length corresponding
to the carrier concentration at the contact.

Let us discuss briefly the steady-state spatial distributions of the quantities
of interest inside the sample. In a general case the carrier concentration is
nonuniform, having maximum values at the contacts due to the electron injection
and decaying toward the middle of the sample.
Accordingly, without an external voltage bias the potential distribution has a
minimum in the middle of the sample due to the space charge.
When a positive voltage is applied to the anode, the minimum is displaced toward
the cathode, while its amplitude tends to diminish.
This minimum provides a potential barrier for the electrons moving between the
contacts, so that a part of the electrons, not having enough energy to go over
the barrier, are reflected back to the contacts.
The most important fact is that the transmission through the barrier is {\em
current dependent}, which is crucial in calculating the noise characteristics.
In the structure the current is limited by the space charge and increases
linearly with
the applied voltage up to a certain value of the external bias when the barrier
vanishes so that all the electrons emitted from the cathode can reach the anode.
Under the latter regime the current is saturated and becomes independent of the
bias.\cite{remark} It is important to stress that in our approach we do not
impose a fixed
number of electrons $N$ to be present inside the sample.
The value of $N$ is determined by the emission rates of the contacts and the
applied bias. Therefore, $N$ fluctuates in time and we can evaluate both the
time-averaged value $\langle N \rangle$ and its fluctuations by means of the
Monte Carlo algorithm.
One can observe that $\langle N\rangle$ is constant at the increasing part of
the current-voltage characteristic, and it decreases with the bias once the
current is saturated.

Under a fixed applied voltage the current density in the structure is given by
$I(t)=(q/L)\sum_{i=1}^{N(t)}v_i(t)$, where $v_i(t)$ is the instantaneous
velocity component along the field direction of the $i$th particle. \cite{gonz}
The current autocorrelation function
$C_I(t)=\langle \delta I(t')\delta I(t'+t)\rangle$ is evaluated from the
sequence $I(t)$ obtained from the Monte Carlo simulation, where the current
fluctuation is given by $\delta I(t)=I(t)-\langle I\rangle$.
To clarify the role of different contributions to the current noise we decompose
the current autocorrelation function into three main contributions
$C_I(t)=C_V(t)+C_N(t)+C_{VN}(t)$ associated, respectively, with the fluctuations
in the mean velocity of electrons $C_V$, the fluctuations in the carrier number
$C_N$, and the velocity-number cross correlation $C_{VN}$.
The corresponding formulas are given by \cite{gonz}

\begin{mathletters}\label{con}
\begin{eqnarray}
C_V(t)&=&\frac{q^2}{L^2} \langle N\rangle^2 \langle \delta v(t') \delta
v(t'+t)\rangle \\
C_N(t)&=&\frac{q^2}{L^2} \langle v\rangle^2 \langle \delta N(t') \delta
N(t'+t)\rangle \\
C_{VN}(t)&=&\frac{q^2}{L^2} \langle v\rangle\langle N\rangle \langle \delta v(t')
\delta N(t'+t) \nonumber \\
&&\phantom{xxxxxx} +  \delta N(t') \delta v(t'+t)\rangle
\end{eqnarray}\end{mathletters}

Fig.\ 1 shows the low-frequency value of the spectral density of current
fluctuations $S_I=2\int_{-\infty}^{\infty}C_I(t)dt$ normalized to $2qI_s$,
where $I_s=q\Gamma=\frac{1}{2}qn_c v_{\rm th}S$ is the saturation current
(notice that this is the maximum current that a contact may provide). 
This normalization is performed in order to compare the results for different 
injection-rate densities (different contact dopings).
We provide the results for two different simulation schemes.
The first one involves a {\em dynamic} PS, which means that any fluctuation of
space charge appeared due to the random injection from the contacts causes a
redistribution of the potential, which is self-consistently updated by solving
Poisson equation at each time step during the simulation to account for the
fluctuations associated to long-range Coulomb interaction.
In the second scheme we use a {\em static} PS to calculate only the stationary
potential profile, and once the steady state is reached, the PS is switched off,
so that the carriers move in the {\em frozen} non-fluctuating electric field
profile.
We have checked that both schemes give exactly the same steady-state spatial
distributions and average current, but the noise characteristics are different.
Several values of $n_c$ (and therefore several injection-rate densities) have
been considered. As $n_c$ increases, space-charge effects become more and more
significant, the dimensionless parameter $\lambda$ being the indicator of their
importance.

In the static case, by increasing the applied voltage $U$ we always obtain an
excellent coincidence with the well-known formula \cite{ziel86} used to describe
the crossover from thermal to shot noise when carrier correlation play no role
(represented in the figure by dashed lines):

\begin{equation} \label{z}
S_I=2q \, (I^+ + I^-)=2qI \coth(qU/2k_B T)
\end{equation}
where $I=I^+ - I^-$ is the total current flowing through the diode, consisting
of two opposing currents, $I^+=I_s\exp[-qV_m/k_BT]$ in the forward-bias
direction and $I^-=I_s\exp[-q(V_m+U)/k_BT]$ in the opposite direction,
$V_m$ being the potential minimum induced by the space charge, which is
dependent on $U$. This agreement supports the validity of the simulation scheme
used for the calculations.
For $qU\ll k_BT$, $I^+\sim I^-$, thermal noise is dominant and $S_I\approx
4qI_s\exp[-qV_m/k_BT]$.
Therefore, for the lowest value of $\lambda$ (when space charge is negligible
and $V_m\to 0$), $S_I\to 4qI_s$, while as $\lambda$ increases $V_m$ becomes
significant and $S_I$ decreases.
When $qU\gtrsim k_BT$, $I^+\gg I^-$, the transition from thermal noise to shot
noise takes place and $S_I\approx2qI^+$. Finally, for the highest values of
$U$, saturation occurs, $V_m$ vanishes, and $S_I\approx 2qI_s$.

For the lowest values of $\lambda$ no difference between the dynamic and static
cases is obviously detected.
However, for higher $\lambda$, when space-charge effects become significant, the
picture is drastically different for the dynamic case.
Starting from $qU\sim k_BT$ the current noise, instead of increasing, decreases
until the proximity of saturation. Under saturation, the results for both
schemes coincide and full shot noise $2qI_s$ is recovered.
When compared with the static case the noise suppression is stronger for higher
$\lambda$ (more important space-charge effects).

To understand the physical reason for the shot-noise suppression, in Fig.\ 2 
we provide the decomposition of $S_I$, calculated with static and dynamic PS, into
the additive contributions $S_V$, $S_N$, and $S_{VN}$ [Eqs.\ (\ref{con})] for
different applied voltages $U$ and $\lambda$=7.72 ($n_c=2.5\times 10^{16}
{\rm cm}^{-3}$).
The contributions of $S_N$ and $S_{VN}$ to the current noise vanish at
equilibrium ($U\to 0$), since they are proportional to $\langle v \rangle^2$
and $\langle v \rangle\to 0$.
Thus, for small biases ($qU\ll k_BT$) $S_I\sim S_V$, which means that the
current noise is thermal noise associated with velocity fluctuations and is
governed by the Nyquist theorem $S_I\approx 4k_BT G$, with $G=dI/dV|_{V=0}$ the
conductance.
For this case the results for the static [Fig.\ 2(a)] and dynamic [Fig.\ 2(b)]
schemes evidently coincide.
However, starting from $qU\sim k_BT$ the difference becomes drastic. For the
dynamic case the velocity-number correlations, represented by $S_{VN}$, are
negative, while for the static case they are positive.
Furthermore, for the current fluctuations calculated using the self-consistent
potential, $S_N$ and $S_{VN}$ are of opposite sign and compensate for each other, 
so that $S_I$ approximately follows $S_V$ as long as the current is space-charge
limited.
As a consequence, the current noise, which now corresponds to shot noise, is
considerably suppressed below the value $2qI$ given by the static case.
This result reflects the fact that as the carriers move through the active
region, the dynamic fluctuations of the electric field modulate the transmission
through the potential minimum and smooth the current fluctuations imposed by the
random injection at the contacts.
Therefore, the coupling between number and velocity fluctuations induced by the
self-consistent potential fluctuations is the main responsible, through
$S_{VN}$, for the shot-noise suppression. This velocity-number coupling becomes
especially pronounced just before the current saturation ($U\approx 7k_BT/q$),
when the potential minimum is close to vanishing completely ($V_m\to 0$) and the
fluctuations of the potential barrier modulate the transmission of the more
populated states of the injected carriers (the low-velocity states). Under
saturation conditions space-charge effects do not modulate the random injection
(no potential minimum is present) and again both dynamic and static cases
provide the same additive contributions and total noise ($2qI_s$).

Finally, in Fig.\ 3 we present the reduction factor $\gamma'$ defined as the
ratio between $S_I$ as calculated with the dynamic PS ($S_{I-d}$) and as given
by Eq.\ (\ref{z}) neglecting the influence of long-range Coulomb interaction,
$\gamma'=S_{I-d}/[2qI\coth(qU/2k_BT)]$.
In the context of our calculations $\gamma'$ is more appropriate than the
standard suppression factor $\gamma=S_{I-d}/(2qI)$, since it covers both the
thermal and shot-noise range of applied voltages. Here it is observed how the
shot-noise suppression becomes more pronounced as $\lambda$ increases.
For example, for $\lambda$=30.9 it reaches 0.04.
Thus, our self-consistent approach predicts much lower values of the suppression
factor than the previous analytical model of van der Ziel and Bosman, \cite{ziel}
where the dependence of the potential minimum and its position on the applied
voltage was not taken into account.

In principle, the value of the parameter $\gamma'$ in our model has no lower
limit. We observe that it follows asymptotically the behavior
$\gamma'\sim k_BT/qU$ in the range where shot-noise suppression is more
pronounced ($qU\gg k_BT$, $U\ll U_{\rm sat}$).
$\gamma'$ can reach a value as low as desired by appropriate increasing the
sample length and/or the carrier concentration at the contact, provided the
transport remains ballistic.
However, with increasing the device length (or lattice temperature) the
carrier transport actually goes from the ballistic to the diffusive regime, and the
shot-noise suppression is washed out.
Therefore, the maximum suppression factor predicted by our calculations for a
system with a given value of the mean free path $\lambda_p$ would be obtained
approximately at $\lambda_{\rm max}\sim\lambda_p/L_{Dc}$.
Moreover, when the carrier concentration at the contact is increased so that the
electron gas becomes degenerate, statistical (Pauli) correlations between the
carriers appear, which will be additive (in the sense of shot-noise suppression)
to the Coulomb correlations.

It should be emphasized that two essential conditions are necessary for the
strong shot-noise suppression due to long-range Coulomb interaction:
(i) the presence of a potential barrier inside the device which controls the
current, and (ii) the carrier transmission through the barrier should depend on
the current.
This fact is quite general and, therefore, the results obtained in the present
paper extend to much more physical situations.
For example, in the recent experiment by Reznikov {\em et al.} \cite{reznikov}
the shot-noise level measured in a quantum point contact in the pinched-off
regime was found to be unexpectedly low (about one-third and less).
In that regime the transport is controlled by the potential barrier present at
the gates, and the both conditions for the shot-noise suppression mentioned
above are fulfilled.
Hence the results obtained in our calculations strongly support the suggestion
of the authors of the experiment that the origin of the discrepancies between
experimental results and theoretical predictions lies in the disregarding of
Coulomb interaction between electrons passing through the contact.
More precisely, the electron flow considerably modifies the potential
distribution inside the contact, yielding the coupling of velocity-number
fluctuations, and resulting in shot-noise suppression.

In conclusion, we have investigated the influence of long-range Coulomb
interaction on shot-noise suppression in ballistic transport by using an
ensemble Monte Carlo simulator self-consistently coupled with a Poisson solver.
We have found that this suppression is stronger as space-charge effects become
more important, and it can be monitored by a dimensionless parameter $\lambda$.
More than one order of magnitude of shot-noise suppression is predicted.
The main contribution to the suppression is found to originate from the
velocity-number correlations induced by the self-consistent field.

This work has been partially supported by the Comisi\'on Inter\-ministerial de
Ciencia y Tecnolog\'{\i}a through the project TIC95-0652.

\begin{figure}
\caption{
Current-noise spectral density $S_I$ vs applied voltage $U$ calculated by using
static (open symbols) and self-consistent (close symbols, solid line) potentials
for several injection-rate densities $n_c$ (in cm$^{-3}$, with the corresponding
$\lambda$): ($\circ$) 10$^{13}$, $\lambda$=0.15; ($\Diamond$) 2$\times 10^{15}$,
$\lambda$=2.18; ($\bigtriangledown$) 10$^{16}$, $\lambda$=4.88;
($\bigtriangleup$) 2.5$\times 10^{16}$, $\lambda$=7.72; ($\Box$) 10$^{17}$,
$\lambda$=15.45; ($\odot$) 4$\times 10^{17}$, $\lambda$=30.9.
The static case is shown to be nicely described by Eq.\ (2) (dashed line).
The dotted lines represent $2qI$ (marked for each injection-rate density by the
corresponding symbol).
}\label{sup}\end{figure}

\begin{figure}
\caption{
Decomposition of the spectral density of current fluctuations $S_I$ into
velocity, number and velocity-number contributions vs applied voltage for the
case $n_c$=2.5$\times 10^{16} {\rm cm}^{-3}$, $\lambda$=7.72 calculated by using
(a) the static and (b) the dynamic Poisson solver.
}\label{cont}\end{figure}

\begin{figure}
\caption{
Shot noise reduction factor $\gamma'$ vs voltage $U$ for several injection-rate
densities $n_c$ (different values of $\lambda$).
}\label{gamp}\end{figure}


\begin{references}

\bibitem[\ast]{byline}
Present address: Dept. F\'{\i}sica Fonamental, Universitat de Barcelona,
Av. Diagonal 647, E-08028 Barcelona, Spain.

\bibitem{jong}
See, e.g., recent review
M. J. M. de Jong and C. W. J. Beenakker,
cond-mat/9611140.

\bibitem{ziel}
A. van der Ziel and G. Bosman, Phys. Status Solidi A {\bf 73}, K93 (1982).

\bibitem{north}
D. O. North, RCA Review {\bf 4}, 441 (1940); {\it ibid.} {\bf 5}, 106 (1941).

\bibitem{birk}
H. Birk, M. J. M. de Jong, and C. Sch\"onenberger,
Phys. Rev. Lett. {\bf 75}, 1610 (1995).

\bibitem{li}
Y. P. Li, D. C. Tsui, J. J. Heremans, J. A. Simmons, and G. W. Weimann,
Appl. Phys. Lett. {\bf 57}, 774 (1990).

\bibitem{reznikov}
M. Reznikov, M. Heiblum, H. Shtrikman, and D. Mahalu,
Phys. Rev. Lett. {\bf 75}, 3340 (1995).

\bibitem{kumar}
A. Kumar, L. Saminadayar, D. C. Glattli, Y. Jin, and B. Etienne,
Phys. Rev. Lett. {\bf 76}, 2778 (1996).

\bibitem{remark}
This behavior contrasts with the exponential and Child-Langmuir $I-V$ dependences
typical of vacuum diodes. It is due to the presence of the opposite current
injected at the anode.

\bibitem{gonz}
T. Gonz\'alez and D. Pardo, J. Appl. Phys. {\bf 73}, 7453 (1993).

\bibitem{ziel86}
A. van der Ziel, {\it Noise in Solid State Devices and Circuits},
(Wiley, New York, 1986).

\end{references}
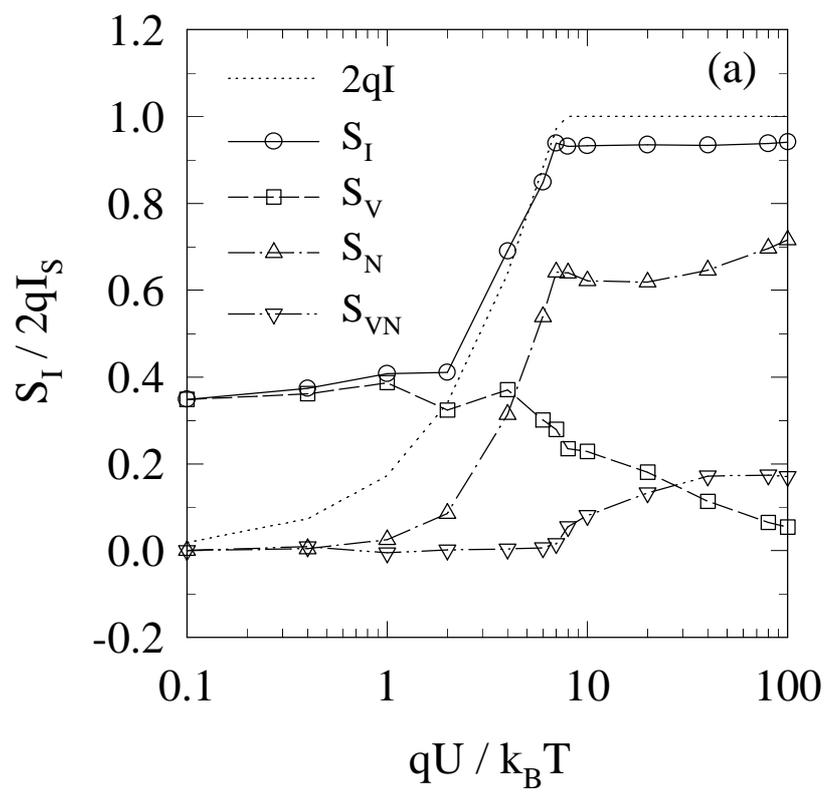
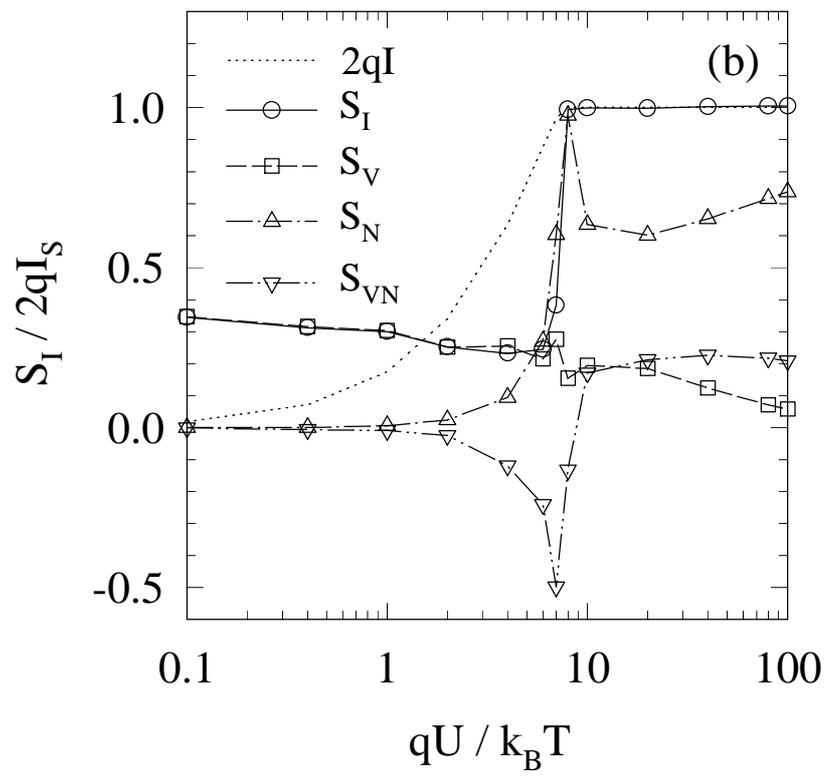
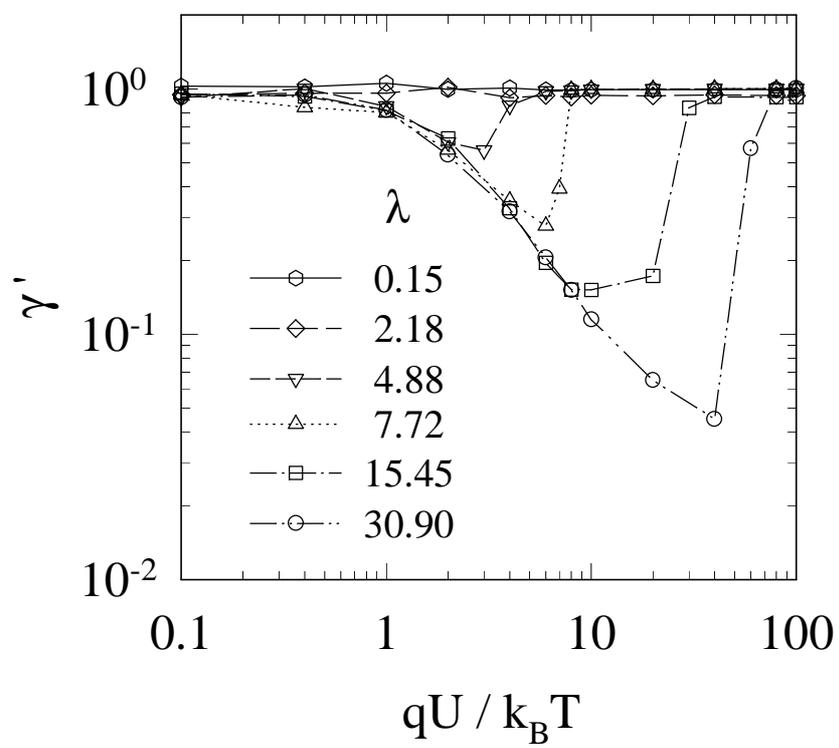
\end{document}